\documentclass[aps,preprint]{revtex4}
\usepackage{amssymb,epsf}
\usepackage{amsmath,amsfonts}
\usepackage{graphicx}
\usepackage{epsfig}
\usepackage{epstopdf}
\usepackage{graphicx}

\begin{document}

\author{N. Farhangkhah \footnote{email address:
farhangkhah@iaushiraz.ac.ir}}
\title{New solutions of exotic charged black holes and
their stability}
\affiliation{Department of Physics, Shiraz Branch, Islamic Azad University, Shiraz 71993,
Iran}

\begin{abstract}
We find a class of charged black hole solutions in third order Lovelock
Gravity. To obtain this class of solutions, we are not confined to the usual
assumption of maximal symmetry on the horizon and will consider the solution
whose boundary is Einstein space with supplementary conditions on its Weyl
tensor. The Weyl tensor of such exotic horizons exposes two charge-like
parameter to the solution. These parameters in addition with the electric
charge, cause different features in compare with the charged solution with
constant-curvature horizon. For this class of asymptotically (A)dS
solutions, the electric charge dominates the behavior of the metric as $r$
goes to zero, and thus the central singularity is always timelike. We also
compute the thermodynamic quantities for these solutions and will show that
the first law of thermodynamics is satisfied. We also show that the extreme
black holes with nonconstant-curvature horizons whose Ricci scalar are zero
or a positive constant could exist depending on the value of the electric
charge and charged-like parameters. Finally, we investigate the stability of
the black holes by analyzing the behavior of free energy and heat capacity
specially in the limits of small and large horizon radius. We will show that
in contrast with charged solution with constant-curvature horizon, a phase
transition occurs between very small and small black holes from a stable
phase to an unstable one, while the large black holes show stability to both
perturbative and non-perturbative fluctuations.
\end{abstract}

\maketitle

\section{Introduction}

One of the predictions of the braneworld with large extra-dimensions is the
possibility of higher-dimensional black hole. A reason for constructing
higher-dimensional theories of gravity is that they provide a framework for
unifying gravity with other interactions. String theory as one important
candidate for such a unified theory, also predicts higher-curvature
corrections to general relativity in addition to the existence of extra
dimensions \cite{string}. For decades we know that Einstein-Hilbert action
is only an effective gravitational action valid for small curvature or low
energies and should be modified by higher-curvature terms. Lovelock theory
of gravity \cite{Lovelock} is the most appropriate one in which the equation
of motion continues to remain second order avoiding ghosts. Lovelock theory
suggests that in higher than four dimensions, higher order terms have to be
added to the usual Einstein-Hilbert action in order to preserve the unique
properties of general relativity in four dimensions. These higher order
gravity terms, are dimensionally extended Euler Poincar\'{e} densities of
two, four-dimensional and so forth manifolds. Most of the researches on this
subject has been concerned with second-order Lovelock gravity known as
Einstein-Gauss-Bonnet (EGB) gravity, in which terms quadratic in the
curvature are added to the action \cite{Cai}. In third-order Lovelock
gravity, Lagrangian and field equations look complicated, but particular
features arising in the solutions accuses the interests to deal with the
higher curvature terms in this theory. There also exist a large number of
works on introducing and discussing various exact black hole solutions of
third order Lovelock gravity \cite{Sham, Lovelockex1, Lovelockex2,
Lovelockex3, Lovelockex4}. Recently, some works have been extended to
general Lovelock gravity to investigate the solutions and their properties
\cite{Lovelockg1, Lovelockg2, Charm, Lovelockg3}. Most of the researches
have been done to derive solutions with maximally symmetric horizons. A
generalization comes through the consideration of horizons which belong to
the more general class of Einstein spaces. In four dimensions, the first
explicit inhomogenous compact Einstein metric was constructed by Page \cite%
{Page} and then was generalized to higher dimensions \cite{Hashimoto}. After
that Bohm constructed metrics with non-constant curvature on products of
spheres \cite{Bohm} and examples in higher-dimensional spacetimes has been
worked on \cite{Gibbons1, LuPa, Gauntlett, Gibbons2, Gibbons3, Gibbons4}.
One of the advantages of considering higher curvature terms in Lovelock
gravity is obtaining new static solution with nonconstant curvature horizon.
In general relativity, substituting the usual $(n-2)$-sphere of the horizon
geometry for an $n$-dimensional spacetime, with an $(n-2)$-dimensional
Einstein manifold will not alter the black hole potential because Einstein's
equations only involve the Ricci tensor. The presence of the Lovelock terms
expose the Riemann curvature tensor to the equations and the new solution
seems to be obtained due to the appearance of the Wyle tensor in the
relation for the Riemann tensor of an Einstein space. In \cite{Dotti}, the
authors considered a static spacetime with generic Einstein space as
dimensional subspace and found that only horizons satisfying the appropriate
conditions on $C_{acde}C_{bcde}$ are allowed, where $C_{abcd}$ is the Weyl
tensor. This constraint appears in the metric and changes the properties of
the spacetime. Various features of such black holes with nontrivial
boundaries like uniqueness and stability in EGB gravity were studied by
several authors \cite{Maeda, Maeda2, Oliva, Tronc}. The Birkhoff`s
theorem in six-dimensional EGB gravity for the case of nonconstant-curvature
horizons with various features has been investigated in \cite{Charm1}. Also
the Birkhoff's theorem is extended in Lovelock gravity for arbitrary base
manifolds using an elementary method \cite{Sourya}. In \cite{Farhang} it is
shown that appearing higher-curvature terms in third order Lovelock gravity,
causes novel changes in the properties of the spacetime with nonconstant
curvature manifold. Some exact solutions with these kinds of
manifolds in Lovelock theory are presented in \cite{Ohashi, Dadh}. In this
article we consider Einstein manifolds of nonconstant curvature and will
investigate charged solution and its properties. In Lovelock theory with $%
U(1)$ field, a charged black hole solution is known \cite{Garraffo}. Also
classes of nonlinear electrodynamics in Einstein and higher derivative
gravity have been studied in \cite{Sheykhi, Sheykhi2, Hendi1}. For charged
black holes with maximally symmetric horizons like spherical or topological
black holes, stability analysis have been performed \cite{Dotti2, Neupane,
Beroiz, Takahashi}. In \cite{Farhang}, it is shown that uncharged black
holes with nonconstant-curvature horizons have unstable phases. Our purpose
is examining the response of instability of such black holes to the charge.

The paper will proceed as follows. In the next section we review the basic
elements of Lovelock gravity and obtain the solution for Lovelock-Maxwell-
system with nonconstant curvature horizon making use of the expressions in
warped geometry for our spacetime ansatz. Also the asymptotic behaviors of
the solution will be discussed. In Sec. \ref{The} the expressions for the
mass, temperature, entropy and electric potential of the solution are
calculated. The stability analysis is also presented by calculating the free
energy and heat capacity in small and large black hole limits for Ricci flat
black holes in which we predict to encounter new features. Finally, we give
some concluding remarks.

\section{\protect\bigskip Charged Solution with Nonconstant-Curvature Horizon%
}

We begin with the action of third order Lovelock gravity in the presence of
electromagnetic field, which is written as
\begin{equation}
I=\int_{\mathcal{M}}d^{n}x\sqrt{-g}\left( 2\Lambda +\mathcal{L}^{(1)}+\alpha
_{2}\mathcal{L}^{(2)}+\alpha _{3}\mathcal{L}^{(3)}-F_{\mu \nu }F^{\mu \nu
}\right) .  \label{Act}
\end{equation}%
where $\Lambda $ is the cosmological constant and $\alpha _{2}$ and $\alpha
_{3}$\ are second and third order Lovelock coefficients and the Maxwell
field strength, or the Faraday tensor, is given by $F_{\mu \nu }:=\partial
_{\mu }A_{\nu }-\partial _{\nu }A_{\mu }$, where $A_{\mu }$ is the vector
potential. The Einstein term $\mathcal{L}^{(1)}$ equals to $R$ and the
second order Lovelock term is $\mathcal{L}^{(2)}=R_{\mu \nu \gamma \delta
}R^{\mu \nu \gamma \delta }-4R_{\mu \nu }R^{\mu \nu }+R^{2}.$ Also $\mathcal{%
L}^{(3)}$ is the third order Lovelock Lagrangian which is described as
\begin{eqnarray}
\mathcal{L}^{(3)} &=&2R^{\mu \nu \sigma \kappa }R_{\sigma \kappa \rho \tau
}R_{\phantom{\rho \tau }{\mu \nu }}^{\rho \tau }+8R_{\phantom{\mu
\nu}{\sigma \rho}}^{\mu \nu }R_{\phantom {\sigma \kappa} {\nu \tau}}^{\sigma
\kappa }R_{\phantom{\rho \tau}{ \mu \kappa}}^{\rho \tau }+24R^{\mu \nu
\sigma \kappa }R_{\sigma \kappa \nu \rho }R_{\phantom{\rho}{\mu}}^{\rho }
\notag \\
&&+3RR^{\mu \nu \sigma \kappa }R_{\sigma \kappa \mu \nu }+24R^{\mu \nu
\sigma \kappa }R_{\sigma \mu }R_{\kappa \nu }+16R^{\mu \nu }R_{\nu \sigma
}R_{\phantom{\sigma}{\mu}}^{\sigma }-12RR^{\mu \nu }R_{\mu \nu }+R^{3}.
\label{ToL}
\end{eqnarray}
The third Lovelock term in eq. (\ref{Act}) has no contribution to the field
equations in six or less dimensional spacetimes, we therefore consider
n-dimensional spacetimes with $n>6$. The gravitational equations following
from the variation of the action (\ref{Act}) with respect to $g_{\mu \nu }$
reads
\begin{equation}
\mathcal{G}_{\mu \nu }:=-\Lambda g_{\mu \nu }+G_{\mu \nu
}^{(1)}+\sum_{p=2}^{3}\alpha _{i}\left( H_{\mu \nu }^{(p)}-\frac{1}{2}g_{\mu
\nu }\mathcal{L}^{(p)}\right) =\kappa _{n}^{2}T_{\mu \nu },  \label{Geq}
\end{equation}%
where%
\begin{equation}
H_{\mu \nu }^{(2)}:=2(R_{\mu \sigma \kappa \tau }R_{\nu }^{\phantom{\nu}%
\sigma \kappa \tau }-2R_{\mu \rho \nu \sigma }R^{\rho \sigma }-2R_{\mu
\sigma }R_{\phantom{\sigma}\nu }^{\sigma }+RR_{\mu \nu }),  \label{Love2}
\end{equation}%
\begin{eqnarray}
H_{\mu \nu }^{(3)} &:&=-3(4R^{\tau \rho \sigma \kappa }R_{\sigma \kappa
\lambda \rho }R_{\phantom{\lambda }{\nu \tau \mu}}^{\lambda }-8R_{%
\phantom{\tau \rho}{\lambda \sigma}}^{\tau \rho }R_{\phantom{\sigma
\kappa}{\tau \mu}}^{\sigma \kappa }R_{\phantom{\lambda }{\nu \rho \kappa}%
}^{\lambda }+2R_{\nu }^{\phantom{\nu}{\tau \sigma \kappa}}R_{\sigma \kappa
\lambda \rho }R_{\phantom{\lambda \rho}{\tau \mu}}^{\lambda \rho }  \notag \\
&&-R^{\tau \rho \sigma \kappa }R_{\sigma \kappa \tau \rho }R_{\nu \mu }+8R_{%
\phantom{\tau}{\nu \sigma \rho}}^{\tau }R_{\phantom{\sigma \kappa}{\tau \mu}%
}^{\sigma \kappa }R_{\phantom{\rho}\kappa }^{\rho }+8R_{\phantom
{\sigma}{\nu \tau \kappa}}^{\sigma }R_{\phantom {\tau \rho}{\sigma \mu}%
}^{\tau \rho }R_{\phantom{\kappa}{\rho}}^{\kappa }  \notag \\
&&+4R_{\nu }^{\phantom{\nu}{\tau \sigma \kappa}}R_{\sigma \kappa \mu \rho
}R_{\phantom{\rho}{\tau}}^{\rho }-4R_{\nu }^{\phantom{\nu}{\tau \sigma
\kappa }}R_{\sigma \kappa \tau \rho }R_{\phantom{\rho}{\mu}}^{\rho
}+4R^{\tau \rho \sigma \kappa }R_{\sigma \kappa \tau \mu }R_{\nu \rho
}+2RR_{\nu }^{\phantom{\nu}{\kappa \tau \rho}}R_{\tau \rho \kappa \mu }
\notag \\
&&+8R_{\phantom{\tau}{\nu \mu \rho }}^{\tau }R_{\phantom{\rho}{\sigma}%
}^{\rho }R_{\phantom{\sigma}{\tau}}^{\sigma }-8R_{\phantom{\sigma}{\nu \tau
\rho }}^{\sigma }R_{\phantom{\tau}{\sigma}}^{\tau }R_{\mu }^{\rho }-8R_{%
\phantom{\tau }{\sigma \mu}}^{\tau \rho }R_{\phantom{\sigma}{\tau }}^{\sigma
}R_{\nu \rho }  \notag \\
&&-4RR_{\phantom{\tau}{\nu \mu \rho }}^{\tau }R_{\phantom{\rho}\tau }^{\rho
}+4R^{\tau \rho }R_{\rho \tau }R_{\nu \mu }-8R_{\phantom{\tau}{\nu}}^{\tau
}R_{\tau \rho }R_{\phantom{\rho}{\mu}}^{\rho }+4RR_{\nu \rho }R_{%
\phantom{\rho}{\mu }}^{\rho }-R^{2}R_{\mu \nu }),  \label{Love3}
\end{eqnarray}
\bigskip and the energy-momentum tensor $T_{\mu \nu }$ is given by%
\begin{equation}
T_{\mu \nu }=2F_{\mu }^{\rho }F_{\rho \nu }-\frac{1}{2}F_{\rho \sigma
}F^{\rho \sigma }g_{\mu \nu }.  \label{EMT}
\end{equation}
Furthermore, variation of the action (\ref{Act}) with respect to the
electromagnetic field reads

\begin{equation}
\nabla _{\nu }F^{\mu \nu }=0  \label{gradF}
\end{equation}

Let us consider the following metric%
\begin{equation}
g_{\mu \nu }dx^{\mu }dx^{\nu }=g_{ab}(y)dy^{a}dy^{b}+r^{2}(y)\gamma
_{ij}(z)dz^{i}dz^{j},  \label{metric1}
\end{equation}%
to be a warped product of a $2$-dimensional \textit{Riemannian} submanifold $%
\mathcal{M}^{2}$ and an $(n-2)$-dimensional submanifold $\mathcal{K}^{(n-2)}$%
. In (\ref{metric1}) $a,b=0,1$ and $i,j$ go from $2,...,n-1$. For a
spherically symmetric spacetime, the metric of $\mathcal{M}^{2}$ is

\begin{equation}
ds^{2}=-f(r)dt^{2}+\frac{1}{f(r)}dr^{2}.  \label{metric2}
\end{equation}%
We assume the submanifold $\mathcal{K}^{(n-2)}$ with the unit metric $\gamma
_{ij}$ to be an Einstein manifold with nonconstant curvature but having a
constant Ricci scalar being

\begin{equation}
\widetilde{R}=\kappa (n-2)(n-3),\text{ \ }  \label{Ricci Sca}
\end{equation}%
with $\kappa $ being the sectional curvature. Hereafter we use tilde for the
tensor components of \ the submanifold \ $\mathcal{K}^{(n-2)}.$ The Ricci
and Riemann tensors of the Einstein manifold are

\begin{eqnarray}
\text{\ \ \ \ \ \ \ }\widetilde{R}_{ij} &=&\kappa (n-3)\gamma _{ij},
\label{Ricci Ten} \\
\widetilde{{R}}{_{ij}}^{kl} &=&\widetilde{{C}}{_{ij}}^{kl}+\kappa ({\delta
_{i}}^{k}{\delta _{j}}^{l}-{\delta _{i}}^{l}{\delta _{j}}^{k})\text{\ },
\label{Riemm Ten}
\end{eqnarray}%
where $\widetilde{{C}}{_{ij}}^{kl}$ is the Weyl tensor of $\mathcal{K}%
^{(n-2)}$.

For the metric (\ref{metric1}) to be a solution of field equations in third
order Lovelock theory in vacuum, it would suffice that the Weyl tensor of
the horizon satisfies the following constraints
\begin{equation}
\sum_{kln}\widetilde{{C}}{_{ki}}^{nl}\widetilde{{C}}{_{nl}}^{kj}=\frac{1}{n}{%
\delta _{i}}^{j}\sum_{kmpq}\widetilde{{C}}{_{km}}^{pq}\widetilde{{C}}{_{pq}}%
^{km}\equiv \eta _{2}{\delta _{i}}^{j},  \label{eta2}
\end{equation}

\begin{eqnarray}
&&\sum_{klnmp}2(4\widetilde{{C}}{^{nm}}_{pk}\widetilde{{C}}{^{kl}}_{ni}%
\widetilde{{C}}{^{pj}}_{ml}-\widetilde{{C}}{^{pm}}_{ni}\widetilde{C}^{jnkl}%
\widetilde{C}_{klpm})  \notag \\
&=&\frac{2}{n}{\delta _{i}}^{j}\sum_{klmpqr}\left( 4\widetilde{{C}}{^{qm}}%
_{pk}\widetilde{{C}}{^{kl}}_{qr}\widetilde{{C}}{^{pr}}_{ml}-\widetilde{{C}}{%
^{pm}}_{qr}\widetilde{C}^{rqkl}\widetilde{C}_{klpm}\right)  \notag \\
&\equiv &\eta _{3}{\delta _{i}}^{j}.  \label{eta3}
\end{eqnarray}
The first constraint was originally introduced by Dotti and Gleiser in \cite%
{Dotti} and the second one which is dictated by the third order Lovelock
term, is obtained in \cite{Farhang}.

In some sense $\eta _{2}$ and $\eta _{3}$ are thought to be topological
charges. We are looking for the charged solutions with nonconstant curvature
horizon, thus we consider the vector potential of the form%
\begin{equation}
A_{\mu }dx^{\mu }=A_{a}(y)dy^{a}=\frac{q}{(n-3)r^{n-3}}dt  \label{vpotential}
\end{equation}%
where $q$ is an arbitrary real constant which is related to the charge of
the solution. With this assumption, equation\ (\ref{gradF}) is trivially
satisfied.\bigskip

Making use of the expressions in warped geometry, the $tt$ component of
field equation (\ref{Geq}) is calculated to be

\begin{eqnarray}
&&\frac{(n-2)}{2r^{6}}\{[r^{5}+2\widehat{\alpha }_{2}r^{3}(\kappa -f)+3%
\widehat{\alpha }_{3}r(\widehat{\eta }_{2}+(\kappa -f)^{2})]f^{^{\prime
}}-(\kappa -f)[(n-3)r^{4}  \notag \\
&&+(n-5)\widehat{\alpha }_{2}r^{2}(\kappa -f)+(n-7)\widehat{\alpha }_{3}(3%
\widehat{\eta }_{2}+(\kappa -f)^{2})]  \notag \\
&&-\left( (n-1)\widehat{\alpha }_{0}+\frac{(n-5)\widehat{\alpha }_{2}%
\widehat{\eta }_{2}}{r^{4}}+\frac{(n-7)\widehat{\alpha }_{3}\widehat{\eta }%
_{3}}{r^{6}}\right) r^{6}\}={\mathcal{G}_{t}}^{t}=-q^{2}r^{10-2n}
\label{Gtt}
\end{eqnarray}%
We define $\widehat{\alpha }_{0}=-2\Lambda /(n-1)(n-2)$, $\widehat{\alpha }%
_{2}=\frac{(n-3)!\alpha _{2}}{(n-5)!}$, $\widehat{\alpha }_{3}=\frac{%
(n-3)!\alpha _{3}}{(n-7)!}$, $\widehat{\eta }_{2}=\frac{(n-6)!\eta _{2}}{%
(n-2)!}$\ and $\widehat{\eta }_{3}=\frac{(n-8)!\eta _{3}}{(n-2)!}$ for
simplicity. We consider $\alpha _{2}$ and $\alpha _{3}$ as positive
parameters. It is also notable to mention that $\widehat{\eta }_{2}$ is
always positive, but $\widehat{\eta }_{3}$ can be positive or negative
relating to the metric of the spacetime. For example cross product
of $p$ $(p\geq 3)$ 2-spheres are Einstein spaces satisfying
conditions (\ref{eta2}) and (\ref{eta3}) having positive $\widehat{\eta }_{3}$
and that of 2-hyperbolas having negative $\widehat{%
\eta }_{3}$. In general if $K^{p}$ denotes a $p$-dimensional maximally symmetric space,
the $q$th products
of such spaces also satisfy conditions (\ref{eta2}) and (\ref{eta3}). Other
non-trivial examples are the complex projective spaces like the standard
Fubini-Study metric or the Bergman space which are considered in \cite%
{Charm1} in six dimensions. The interesting point of such black holes is that
they can lead via Kaluza-Klein compactification to
lower-dimensional scalar-tensor black holes \cite{Charm2, Karel}. See also
Ref. \cite{Ohashi} for more examples of Einstein spaces. Introducing

\begin{equation}
\psi (r)=\frac{\kappa -f(r)}{r^{2}},  \label{Psi}
\end{equation}
and integrating $\int r^{n-2}\mathcal{G}_{t}^{t}dr$, one obtains

\begin{equation}
\left( 1+\frac{3\widehat{\alpha }_{3}\widehat{\eta }_{2}}{r^{4}}\right) \psi
+\widehat{\alpha }_{2}\psi ^{2}+\widehat{\alpha }_{3}\psi ^{3}+\widehat{%
\alpha }_{0}+\frac{\widehat{\alpha }_{2}\widehat{\eta }_{2}}{r^{4}}+\frac{%
\widehat{\alpha }_{3}\widehat{\eta }_{3}}{r^{6}}-\frac{m}{r^{n-1}}+\frac{%
2q^{2}}{(n-2)(n-3)r^{2(n-2)}}=0,  \label{Eq3}
\end{equation}%
where $m$ is the integration constant.

This cubic equation can admit three real roots. One of the real
solutions to this equation may be written as:
\begin{eqnarray}
\psi (r) &=&-\frac{\alpha _{2}r^{2}}{3\widehat{\alpha }_{3}}\left\{ 1-\left(
j(r)\pm \sqrt{\gamma +j^{2}(r)}\right) ^{1/3}+\gamma ^{1/3}\left( j(r)\pm
\sqrt{\gamma +j^{2}(r)}\right) ^{-1/3}\right\} ,  \notag \\
j(r) &=&-1+\frac{9\widehat{\alpha }_{3}}{2\widehat{\alpha }_{2}^{2}}-\frac{27%
\widehat{\alpha }_{3}^{2}}{2\alpha _{2}^{3}}\left( \widehat{\alpha }_{0}-%
\frac{m}{r^{n-1}}+\frac{\widehat{\alpha }_{3}\widehat{\eta }_{3}}{r^{6}}+%
\frac{q^{2}}{(n-3)r^{2(n-2)}}\right) ,\text{ \ }  \notag \\
\text{\ \ \ \ }\gamma  &=&\left( -1+\frac{3\widehat{\alpha }_{3}}{\widehat{%
\alpha }_{2}^{2}}+\frac{9\widehat{\alpha }_{3}^{2}\widehat{\eta }_{2}}{%
\widehat{\alpha }_{2}^{2}r^{4}}\right) ^{3},  \label{fstat}
\end{eqnarray}%
One may note that since the constant $\widehat{\eta }_{2}$
and $\widehat{\eta }_{3}$ are evaluating on
the $(n-2)$-dimensional boundary, $\widehat{\eta }%
_{3}$ appears in the above equation only for $n\geq 8$.
Thus in order to have the effects of non-constancy of the
curvature of the horizon in third order Lovelock gravity, $n$ should be larger
than seven. One may note that solution (\ref{fstat}) reduces to the
algebraic equation of Lovelock gravity for charged solution with constant
curvature horizon when $\widehat{\eta }_{2}=\widehat{\eta }_{3}=0$.

Here we pause to add some comments on the asymptotic behavior of the
solution. The behavior of the metric function $f$ around $r\rightarrow
\infty $ is exactly the same as the uncharged solution, because the term
including charge vanishes at infinity. Using Eq. (\ref{Psi}) and
taking the $r\rightarrow \infty $ limit of Eq. (\ref{Eq3}), one obtains
\begin{equation}
(k-f_{\infty })r^{4}+\hat{\alpha}_{2}[(k-f_{\infty })^{2}+\hat{\eta}%
_{2}]r^{2}+\hat{\alpha}_{3}[(k-f_{\infty })^{3}+3\hat{\eta} _{2}(k-f_{\infty})+\hat{\eta}%
_{3}]+\hat{\alpha}_{0}=0,  \label{psiinf}
\end{equation}%
where the constant $f_{\infty }$ is the value of $f$ at infinity.
The asymptotic AdS solution exists if Eq. (\ref{psiinf}) has positive real
roots. One may note that in the case $\hat{\eta}_{2}=\hat{\eta}_{3}=0$ and $%
k=1$, one of the roots of Eq. (\ref{psiinf}) for $\widehat{\alpha }%
_{0}=0$ will be $f_{\infty }=1$. That is, the third order Lovelock gravity
with spherical horizon can be asymptotically flat \cite{Sham}. See \cite%
{Myers, Vah} for more details on the asymptotic behavior.

The Kretschmann scalar $R_{\mu \nu \rho \sigma }R^{\mu \nu \rho \sigma }$
diverges at $r=0$. Hence, there is an essential singularity located at $r=0$%
. The dominant term of the metric function around $r=0$ regarding Eq. (\ref%
{Eq3}) is
\begin{equation}
f(r)\simeq (\frac{2q^{2}}{(n-2)(n-3)\widehat{\alpha }_{3}r^{2n-10}})^{1/3}
\end{equation}%
As one can see from the above equation the dominant term at $r=0$ is the
charge term and therefore the central singularity is always timelike, in
contrast with the uncharged solution which possesses a spacelike central
singularity.

\begin{figure}[tbp]
\centering {\includegraphics[width=7cm]{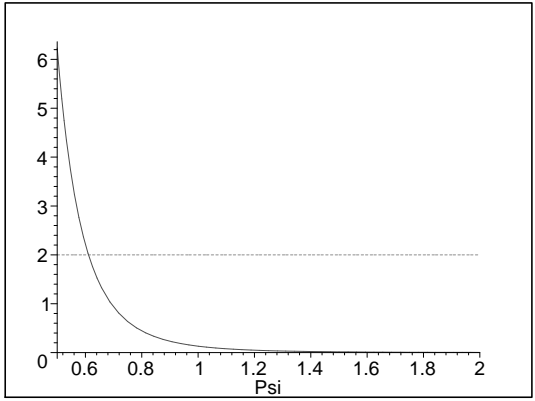}}
\caption{$C(\protect\psi )$ (line) and $q^{2}$ (dotted) versus $\protect\psi
$ for $n=8$, $\hat{\protect\alpha}_{0}=1$, $\hat{\protect\alpha}_{2}=0.2$, $%
\hat{\protect\alpha}_{3}=0.05$, $\hat{\protect\eta}_{2}=0.5$ and $\hat{%
\protect\eta}_{3}=0.006,q=2$.}
\label{psi1}
\end{figure}

\begin{figure}[tbp]
\centering {\includegraphics[width=7cm]{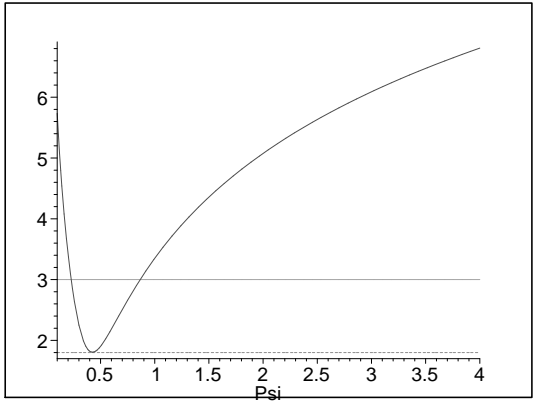}}
\caption{$B(\protect\psi )$ (line) and $m_{ext}$ (dotted) versus $\protect%
\psi $ for $n=8$, $\hat{\protect\alpha}_{0}=1$, $\hat{\protect\alpha}_{2}=0.2
$, $\hat{\protect\alpha}_{3}=0.05$, $\hat{\protect\eta}_{2}=0.5$ and $\hat{%
\protect\eta}_{3}=0.006,q=2$. It can be seen that there exist horizons if $%
m\geq m_{ext}$.}
\label{psi2}
\end{figure}

As it is known, Killing horizon for a black hole is defined by $f(r_{h})=0.$
In order to investigate the existence of the horizon of black hole we
consider Eq. (\ref{Eq3}) for the probable existing $r_{h}$. From the
definition of $\psi $ in Eq. (\ref{Psi}), for $\kappa \neq 0$ we substitute $%
r_{h}=(\kappa /\psi _{h})^{1/2}$ to obtain
\begin{eqnarray}
&&m(\frac{\psi _{h}}{\kappa })^{\frac{n-1}{2}}-\frac{2q^{2}}{(n-2)(n-3)}(%
\frac{\psi _{h}}{\kappa })^{n-2}  \notag \\
&=&\widehat{\alpha }_{0}+\psi _{h}+\widehat{\alpha }_{2}(\kappa ^{2}+%
\widehat{\eta }_{2})(\frac{\psi _{h}}{\kappa })^{2}+\widehat{\alpha }%
_{3}(\kappa ^{3}+3\widehat{\eta }_{2}\kappa +\widehat{\eta }_{3})(\frac{\psi
_{h}}{\kappa })^{3}\equiv A(\psi _{h})  \label{APsi}
\end{eqnarray}%
Solving this equation for $m$ we get%
\begin{equation}
m=\frac{2q^{2}}{(n-2)(n-3)}(\frac{\psi _{h}}{\kappa })^{\frac{n-3}{2}%
}+A(\psi _{h})(\frac{\psi _{h}}{\kappa })^{-\frac{n-1}{2}}\equiv B(\psi _{h})
\label{BPsi}
\end{equation}%
The solutions of this equation give the horizon radius. Solving $\partial _{\psi
}B(\psi )=0$, we obtain
\begin{eqnarray}
q^{2} &=&\frac{(n-2)}{2}\{\widehat{\alpha }_{0}(n-1)(\frac{\psi }{\kappa }%
)^{2-n}+\kappa (n-3)(\frac{\psi }{\kappa })^{3-n}+(n-5)\widehat{\alpha }%
_{2}(\kappa ^{2}+\widehat{\eta }_{2})(\frac{\psi }{\kappa })^{4-n} \\
&&+(n-7)\widehat{\alpha }_{3}(\kappa ^{3}+3\widehat{\eta }_{2}\kappa +%
\widehat{\eta }_{3})(\frac{\psi }{\kappa })^{5-n} \\
&\equiv &C(\psi )
\end{eqnarray}%
We plot function $C(\psi )$ versus $\psi .$ The cross point of the curve $%
C(\psi )$ with $q^{2}$ is the real root for the equation $C(\psi )=q^{2}$
that is shown in Fig. (\ref{psi1}). We call it $\psi _{\min }$ for which the
function $B(\psi )$ has its extreme value. Figure (\ref{psi2}) shows $B(\psi
)$ versus $\psi $ with its minimum at $\psi =\psi _{\min }$. The
intersection of the horizontal line $m$ and this curve gives the radius of
the horizon. There exist horizons if $m\geq m_{ext}$ where $m_{ext}$ is
defined as
\begin{equation}
m_{ext}=\frac{q^{2}}{(n-3)}\psi _{\min }^{\frac{n-3}{2}}+A(\psi _{\min
})\psi _{\min }^{-\frac{n-1}{2}}.
\end{equation}%
It is worth noting that the Eq. (\ref{BPsi}%
) may have real roots if we set $m=0$. This fact is due to the existence
of $\widehat{\eta}_{3}$ that could be negative in the equations (\ref{APsi}) and (\ref{BPsi}).
Thus charged black holes with $m=0$
and nonconstant-curvature horizon may have horizon. This does not happen for
the solution with constant-curvature horizon or the solution with
nonconstant-curvature horizon in second order Lovelock theory. Also,
the reader notes that $m_{ext}$ could be negative for negative $\widehat{\eta}_{3}$, and therefore
$m=0$ is larger than $m_{ext}$ which is negative.

\section{Thermodynamics of Black Hole Solutions}

\label{The} The surface gravity on the Killing horizon is $%
(1/2)(df/dr)|_{r=r_{h}}$, from which the temperature of the horizon $T$
could be written as

\begin{equation}
T=\frac{(n-1)r_{h}^{6}\widehat{\alpha }_{0}+(n-3)\kappa r_{h}^{4}+(n-5)%
\widehat{\alpha }_{2}(\widehat{\eta }_{2}+\kappa ^{2})r_{h}^{2}+(n-7)%
\widehat{\alpha }_{3}(\widehat{\eta }_{3}+3\kappa \widehat{\eta }_{2}+\kappa
^{3})-\frac{2q^{2}}{(n-2)}r^{10-2n}}{4\pi r_{h}[r_{h}^{4}+2\kappa \widehat{%
\alpha }_{2}r_{h}^{2}+3\widehat{\alpha }_{3}(\widehat{\eta }_{2}+\kappa
^{2})]},  \label{Temp}
\end{equation}%
where $r_{h}$ is the radius of the outer horizon. On the other hand, the
entropy on the Killing horizon is calculated using the Wald prescription
which is applicable for any black hole solution of which the event horizon
is a killing horizon \cite{Wald}. The Wald entropy is defined by the
following integral performed on $(n-2)$-dimensional spacelike bifurcation
surface

\begin{equation}
S=-2\pi \oint d^{n-2}x\sqrt{h}Y,\text{ \ \ \ \ \ }Y=Y^{abcd}\widehat{%
\varepsilon }_{ab}\widehat{\varepsilon }_{cd},\text{\ \ \ \ \ \ }Y^{abcd}=%
\frac{\partial \mathcal{L}}{\partial R_{abcd}}  \label{entropy}
\end{equation}%
in which $\mathcal{L}$ is the Lagrangian and $\widehat{\varepsilon }_{ab}$
is the binormal to the horizon. As we mentioned before, $\mathcal{L}_{1},$ $%
\mathcal{L}_{2}\mathcal{\ }$and $\mathcal{L}_{3},$ are Einstein,
Gauss-Bonnet and third order Lovelock Lagrangians respectively, from which
we obtain $Y_{1},$ $Y_{2}$ and $Y_{3.}$ Following the given description, $%
Y_{1}$ and $Y_{2}$ \ and $Y_{3}$ are calculated to be

\begin{equation}
Y_{1}=-\frac{1}{8\pi }  \label{Ein-entropy}
\end{equation}

\begin{equation}
Y_{2}=-\frac{\widehat{\alpha }_{2}}{4\pi }%
[R-2(R_{t}^{t}+R_{r}^{r})+2R_{tr}^{tr}]  \label{Gauss-entropy}
\end{equation}

\begin{eqnarray}
Y_{3} &=&-\frac{3\widehat{\alpha }_{3}}{4\pi }\{-12({R^{tm}}_{tn}{R^{rn}}%
_{rm}-{R^{tm}}_{rn}{{R^{r}}_{mt}}^{n})+12R^{trmn}R_{trmn}-24[{R^{tr}}_{tm}{%
R_{r}}^{m}-{R^{tr}}_{rm}{R_{t}}^{m}  \notag \\
&&+\frac{1}{4}(R_{mnpr}R^{mnpr}+R_{mnpt}R^{mnpt})]+3(2R{R^{tr}}_{tr}+\frac{1%
}{2}R_{mnpq}R^{mnpq})  \notag \\
&&+12({R^{t}}_{t}{R^{r}}_{r}-{R^{t}}_{r}{R^{r}}_{t}+{R^{r}}_{mrn}R^{mn}+{%
R^{t}}_{mtn}R^{mn})+12(R^{rm}R_{rm}+R^{tm}R_{tm})  \notag \\
&&-6[R_{mn}R^{mn}+R({R^{r}}_{r}+{R^{t}}_{t})]+\frac{3}{2}R^{2}\}.
\label{Lov-entropy}
\end{eqnarray}%
Substituting in Eq. (\ref{entropy}), and making use of Eq. (\ref{Riemm Ten}%
), one calculates the entropy to be

\begin{equation}
S=-2\pi \{Y_{1}+Y_{2}+Y_{3}\}=\frac{r_{h}^{n-2}}{4}\left\{ 1+\frac{2\kappa
\widehat{\alpha }_{2}(n-2)}{r_{h}^{2}(n-4)}+\frac{3\widehat{\alpha }%
_{3}(n-2)(\widehat{\eta }_{2}+\kappa ^{2})}{r_{h}^{4}(n-6)}\right\} .
\label{Entro}
\end{equation}

The charge of the black holes per unit volume can be found by calculating
the flux of the electric field at infinity, yielding

\begin{equation}
Q=\frac{q}{4\pi }.  \label{charge}
\end{equation}%
The electric potential , measured at infinity with respect to the horizon,
is defined by \cite{Caldarelli}

\begin{equation}
\Phi =A_{\mu }\chi ^{\mu }\mid _{r\rightarrow \infty }-A_{\mu }\chi ^{\mu
}\mid _{r\rightarrow r_{h}}.  \label{Phi1}
\end{equation}%
Using $\chi =\partial /\partial t$ as the null generator of the horizon, one
finds

\begin{equation}
\Phi =\frac{q}{(n-3)r_{h}^{n-3}}.
\end{equation}

Also we obtain the relation for the mass density, from Eqs. (\ref{Eq3}) and (%
\ref{Psi}) , which admits the relation below

\begin{eqnarray}
M &=&\frac{(n-2)m}{16\pi }=\frac{(n-2)}{16\pi }[\widehat{\alpha }%
_{0}r^{n-1}+\kappa r^{n-3}+\widehat{\alpha }_{2}[\kappa^{2}+\widehat{\eta}
_{2}]r^{n-5}+\widehat{\alpha }_{3}[\kappa^{3}+3\widehat{\eta} _{2}\kappa+%
\widehat{\eta}_{3}]r^{n-7}  \notag \\
&&+\frac{2q^{2}}{(n-2)(n-3)r^{n-3}}].  \label{ADM}
\end{eqnarray}%
Making use of Eqs. (\ref{Temp}), (\ref{Entro}), (\ref{charge}) and (\ref{ADM}%
), one may note that the thermodynamic quantities calculated in this section
satisfy the first law of thermodynamics $dM=T\partial {S+\Phi \partial Q}.$

\section{Stability in Canonical Ensemble For The Case of $\protect\kappa=0$}

\label{Stab}

The stability analysis of a thermodynamic system with respect to the small
variations of the thermodynamic coordinates, is performed by analyzing the
behavior of the entropy near equilibrium. The number of thermodynamic
variables depends on the ensemble that is used. In the canonical ensemble,
the charge is fixed, and therefore positive heat capacity, $C=T(\partial
S/\partial T)_{Q}$, implies that the black hole is locally stable. However,
to analyze the global stability, we should check the free energy of the
black hole which is defined by $F:=M-TS,$ whereby negative value ensures
global stability \cite{Haw-Page}. Before investigating the behavior of heat
capacity, we check the behavior of temperature for small and large black
holes. As mentioned before, it is seen from Eq. (\ref{Temp}), that extreme
black hole exists when $T(r_{h\text{ }})=0$, from which we get:%
\begin{equation}
q_{ext}=r^{n-5}\sqrt{\frac{(n-2)}{2}[(n-1)r_{h}^{6}\widehat{\alpha }%
_{0}+(n-5)\widehat{\alpha }_{2}\widehat{\eta }_{2}r_{h}^{2}+(n-7)\widehat{%
\alpha }_{3}\widehat{\eta }_{3}]}.  \label{qext}
\end{equation}

As we mentioned before, $\widehat{\eta }_{3}$ can be positive or negative
relating to the metric of the spacetime. Therefore we consider these two
cases separately:

a) The case $\widehat{\eta }_{3}>0:$

\begin{figure}[tbp]
\centering {\includegraphics[width=7cm]{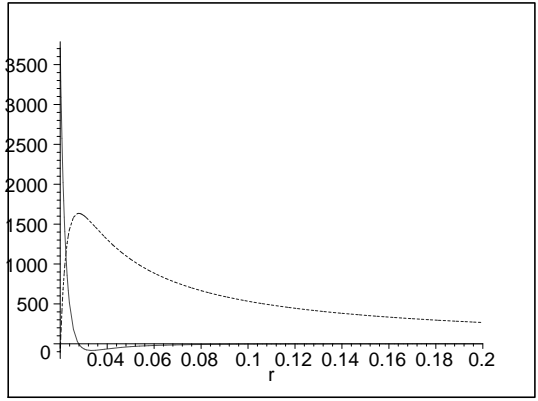}}
\caption{$10^{-3}(\partial T/\partial S)_{Q}$ (line) and $10^{2}T$ (dotted)
versus $r_{h}$ for $n=8 $, , $\hat{\protect\alpha}_{0}=1$, $\hat{\protect%
\alpha}_{2}=0.2$, $\hat{\protect\alpha}_{3}=0.05$, $\hat{\protect\eta}%
_{2}=0.5$ and $\hat{\protect\eta}_{3}=0.006, q=1$.}
\label{lss}
\end{figure}

\begin{figure}[tbp]
\centering {\includegraphics[width=7cm]{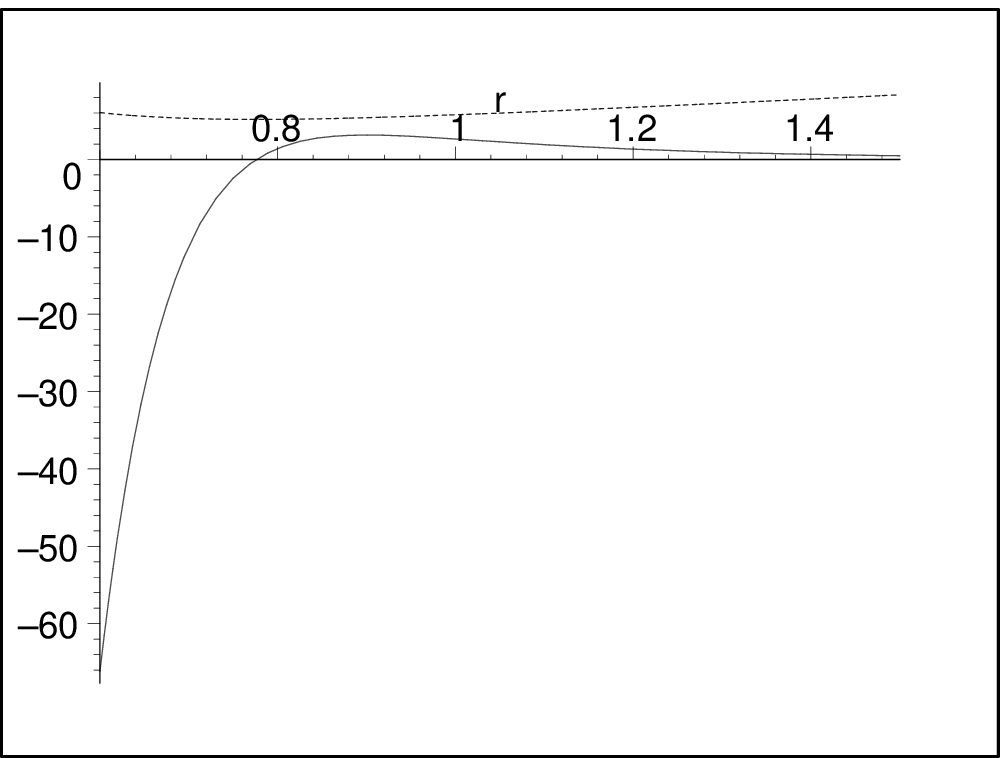}}
\caption{$(\partial T/\partial S)_{Q}$ (line) and $T$ (dotted) versus $r_{h}$
for $n=8$, $\hat{\protect\alpha}_{0}=1$, $\hat{\protect\alpha}_{2}=0.2$, $%
\hat{\protect\alpha}_{3}=0.05$, $\hat{\protect\eta}_2=0.5$ and $\hat{\protect%
\eta}_3=0.006, q=1$.}
\label{lsml}
\end{figure}

\begin{figure}[tbp]
\centering {\includegraphics[width=7cm]{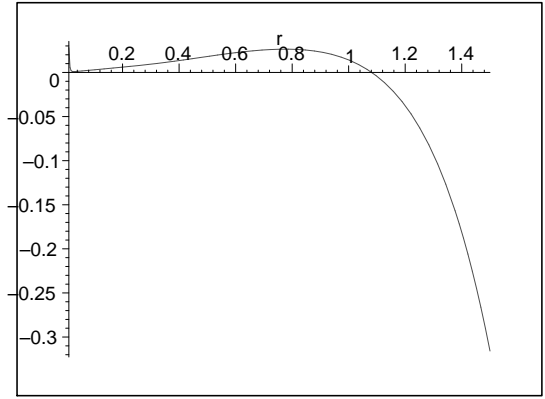}}
\caption{$F$ versus $r_{h}$ for $n=8$, $\hat{\protect\alpha}_{0}=1$, $\hat{%
\protect\alpha}_{2}=0.2$, $\hat{\protect\alpha}_{3}=0.05$, $\hat{\protect\eta%
}_2=0.5$ and $\hat{\protect\eta}_3=0.006, q=1$.}
\label{gs}
\end{figure}

In this case $q_{ext}$ is always real and extreme black hole exists for $%
q=q_{ext}.$ To investigate the stability of the black holes in this case we
check the positivity of $(\partial T/\partial S)_{Q}$ in the regions where $T
$ is positive, because the temperature of a physical black hole is positive.
We plot the curve of $T$ and $(\partial T/\partial S)_{Q}$ versus the radius
for very small black holes in Fig. (\ref{lss}) and for small and large black
holes in figure (\ref{lsml}). It is seen that there is a phase transition
between very small and small black holes from a stable to an unstable phase,
which is characterized by the sign change in heat capacity. This is due to
the existence of charge in the solution and does not occur in uncharged
solution with nonconstant curvature horizon \cite{Farhang}. Large black
holes have positive heat capacity and are locally stable. To perform the
analysis of global stability we depict $F$ versus $r$ in figure (\ref{gs}),
from which we note that small black holes are globally unstable while large
ones are stable.

b) The case $\widehat{\eta }_{3}<0:$

\begin{figure}[tbp]
\centering {\includegraphics[width=7cm]{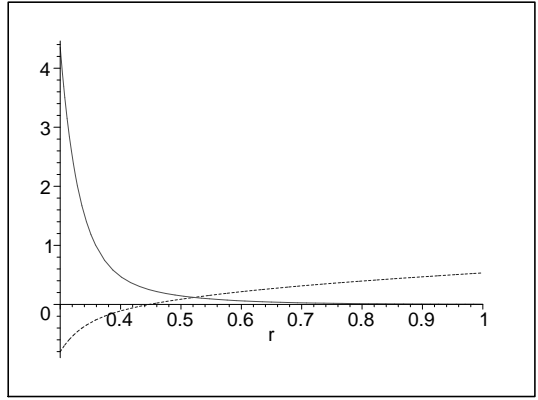}}
\caption{$(\partial T/\partial S)_{Q}$ (line) and $T$ (dotted) versus $r_{h}$
for $n=8$, , $\hat{\protect\alpha}_{0}=1$, $\hat{\protect\alpha}_{2}=0.2$, $%
\hat{\protect\alpha}_{3}=0.05$, $\hat{\protect\eta}_{2}=0.5$ and $\hat{%
\protect\eta}_{3}=-2, q=1$. }
\label{lsme}
\end{figure}

For this case $\widehat{\eta }_{3}$ should satisfy the following condition \
\begin{equation}
\left\vert \widehat{\eta }_{3}\right\vert <\frac{(n-1)r_{h}^{6}\widehat{%
\alpha }_{0}+(n-5)\widehat{\alpha }_{2}\widehat{\eta }_{2}r_{h}^{2}}{(n-7)%
\widehat{\alpha }_{3}}.
\end{equation}%
in order to have a real value for $q_{ext}$. We plot the curve of $T$ and $%
(\partial T/\partial S)_{Q}$ versus $r$ in figure (\ref{lsme}) for $q<q_{ext}
$ as temperature is positive in this case. It is seen that $(\partial
T/\partial S)_{Q}$ is positive in the range that $T$ is positive and
thereby, charged black holes with negative $\widehat{\eta }_{3}$, are
locally stable. The plot of free energy and analysis of global stability in
this case are like the previous case.

\section{Concluding Remarks}

In this study, we presented charged solution of Lovelock gravity with
nonconstant-curvature horizon. We considered a spacetime which is a cross
product of a Lorentzian spacetime and a space with nonconstant-curvature
horizon. With this assumption the usual assumption of maximally symmetry is
relaxed, leading to a new class of black hole solution which introduces
novel chargelike parameter to the black hole potential, in addition to the
electric charge. These parameters are obtained by imposing two conditions on
the Weyl tensor of the Einstein space and appear with the advantage of
higher curvature terms in third order Lovelock equations. It was shown that
the electric charge dominates the behavior of the metric function as $r$
approaching zero, and the central singularity is always timelike, in
contrast with the uncharged solution which possesses a spacelike central
singularity. By investigating the asymptotic behavior of the metric at
infinity, we showed that the solution could be asymptotically AdS in
contrast with the solutions with constant-curvature horizon that could be
flat for $\widehat{\alpha }_{0}=0$ and $\kappa=1$. By introducing the condition for the existence
of the event horizon of the solution, we mentioned that charged black holes with
nonconstant-curvature horizon and $m=0$ could possess event horizon.
This property does not appear for black hole with constant-curvature horizon
or the one with nonconstant-curvature horizon in second order Lovelock theory.
Then we proceeded by calculating the
mass, temperature, entropy and electric potential of the black hole in terms
of horizon radius and the first law of thermodynamics was shown to be hold
for this class of solution. In order to show the differences of charged and
uncharged black holes with nonconstant-curvature horizons, we went through
the properties of black holes with $\kappa =0$. We saw that extreme charged
black hole could exist depending on the values of electric charge $q$ and $%
\widehat{\eta }_{3}$ appearing in the expression for temperature. Also we
mentioned that the entropy of the black hole with $\kappa =0$ does not obey
the area law exactly like that of uncharged solution. To check the stability
of the solutions, confining to the canonical ensemble, we calculated the
heat capacity and free energy in the small and large black hole regimes. The
main difference that occurs due to the existence of charge is that $\kappa =0
$ charged solution with nonconstant-curvature horizon for $\widehat{\eta }%
_{3}>0$ shows a transition between a thermodynamically stable phase for very
small black holes to a thermodynamically unstable phase for small ones which
does not appear for uncharged black holes. Also calculations show that small
charged black holes have positive free energy and are globally unstable
while large black holes possess negative free energy and positive heat
capacity showing their stability to both perturbative and non-perturbative
fluctuations.

\acknowledgments{This work has been supported financially by Research Council of
Shiraz Branch of Islamic Azad University, Iran}

\end{document}